\documentclass[doublecol]{epl2}

\usepackage{color}

\usepackage{graphicx}
\usepackage{bm}
\usepackage{amssymb, dsfont}

\title{Motility-sorting of self-propelled particles in micro-channels}

\author{Andrea~Costanzo, Jens~Elgeti, Thorsten~Auth, Gerhard~Gompper and Marisol~Ripoll}

\institute{Theoretical Soft Matter and Biophysics, Institute of Complex Systems \\ 
Forschungszentrum J\"{u}lich GmbH, D-52425 J\"{u}lich, Germany} 


\abstract{Spontaneous segregation of run-and-tumble particles
  with different velocities in microchannels is investigated by
  numerical simulations. Self-propelled particles are known to
  accumulate in the proximity of walls. Here we show how fast
  particles expel slower ones from the wall leading to a segregated
  state. The mechanism is understood as a function of particle
    velocities, particle density, or channel width.  In the
  presence of an external fluid flow, particles with two different
  velocities segregate due to their different particle fluxes.
  Promising applications can be found in the development of
  microfluidic lab-on-a-chip devices for sorting of particles with
  different motilities.
}

\begin{document}
\maketitle

{\bf Introduction.} - Self-propelled particles are small biological or
synthetic systems that transform energy into directed
motion~\cite{RewVicsek}. Microorganisms like bacteria, sperm cells or
other eukaryotic cells, as well as synthetic particles like nanorods
or Janus colloids, employ various strategies to generate directed
motion in a fluid environment~\cite{lauga-rew,Yang-dimer}. Ensembles
of self-propelled particles can form
clusters~\cite{YangGompperPRE,Wensink-PNAS-2012,Marchetti-PRL-2012,Abkenar13,Wysocki-EPL-2014,Buttinoni,Bocquet,Cates-continuum} and concentrate in
space regions delimited by walls of funnels~\cite{gala,Tailleur2009}.
Furthermore, active particles have been shown to accumulate
in the proximity of the
walls~\cite{Elgeti-EPL2009,Elgeti-EPL2013,LoewenPRE}, resulting for
example in trapping of particles in
microwedges~\cite{LoewenPRL,LoewenPRE2013}, spiral vortex
  formation in circular confinement~\cite{Wioland}, or depletion of
  elongated particles from low-shear regions~\cite{Rusconi}.  
Furteher interesting phenomena, which are only possible with active
particles, are the upstream swimming in microchannels with capillary
flow~\cite{mio-jpcm,ZoettlPRL2012,ZoettlEPJE2013,HillUpstream},
spontaneous directed rotation of asymmetric
wheels~\cite{SimMot,ExpMot,ExpMotSok}, or net particle flux in static
potentials~\cite{epl-mio,MarchesoniPRL2013,Stark-SymmRat-PRE2013}.

Populations of active particles in nature frequently have a wide
distribution of certain properties.  For example, sperm cells have a
range of velocities, and only a few of them are able to reach the ovum
and fertilize it; Janus colloids have a range of sizes and chemical
activity.  Nature has developed strategies to sort particles with
certain properties depending on their functionality. Sorting
strategies are also interesting in technological applications for
self-propelled particles.  Thus, a deeper understanding of sorting
mechanisms is of great fundamental and technological importance, which
has already originated numerous studies.  The famous Maxwell's
demon is able to separate fast and slow gas molecules,
unfortunately by violating the second law of thermodynamics. At the
macroscopic scale, several studies focused on the problem of sorting
particles, exploiting entropic means~\cite{entr-splitter},
gravity~\cite{brasilian-nut}, and microfluidic 
devices~\cite{sort-microflu-brownian,sort-microflu-brownian3,sort-microflu-brownian2,volpe-sm2011,volpe-sm2013}.
At the microscopic scale, spontaneous segregation of active and
passive particles, freely swimming in a two-dimensional box with
periodic boundary conditions, has been numerically
investigated~\cite{McCandlish}. Separation of self-propelled
  particles with different motilities have been investigated for
    asymmetric obstacles~\cite{marconi-pre2013} and
  using centrifuges~\cite{rdl-sm2013}.

In this work, we investigate numerically how the accumulation of
self-propelled particles at walls can be exploited to sort particles
with different velocities. Accumulation of faster particles in the
proximity of the walls is characterized for microchannels as a
function of particle velocities, channel width, and particle
concentration. In the presence of a capillary flow, it is shown that
the downstream flux is enriched with the slower self-propelled
particles.  The combination of a capillary flow with membranes shows
accumulation at both channel ends, with the downstream end enriched in
slow particles, and the upstream end populated almost exclusively by
fast particles. Potential applications in fields as diverse as
microfluidics, biology or even reproductive techniques motivate our
study.

{\bf Model.} - We employ a minimal model for self-propelling
run-and-tumble bacteria~\cite{mio-jpcm}, aimed at mimicking the motion
of {\it E.~coli} \cite{E_coli}.  Each particle is described by two
beads of diameter $a$ rigidly connected at distance $a$, such that the
length of the particle is $l=2a$.  During the run time, the particle
 center of mass is subjected to a force directed along its main axis,
which determines the self-propulsion velocity $v$ via its friction
coefficient in the fluid. The stochastic tumbling event interrupts the
run at random times with constant rate and duration, by imposing a
random torque on the particle which lets a free particle rotate by a
random angle uniformly distributed in $(0,2\pi)$.  We take the
duration of the tumbling $\tau$ as time unit.  Distances are measured
in units of the particle length $l$, such that for example velocities
are given in units of $l/\tau$.  The average time between tumbling
events is chosen to be $10\tau$ and the average
  free-run-length is $\lambda= 10 l$.
Previous work with this
  model~\cite{mio-jpcm}, as well as other recent
  experimental~\cite{drescher11,Buttinoni} and
  simulation~\cite{Elgeti-EPL2009,Elgeti-EPL2013} studies, indicate
  that the role of hydrodynamic interactions in wall or cluster
  aggregation is often negligible in comparison with steric interactions,
  especially for elongated particles.  Hydrodynamic interactions are
  therefore disregarded here. Equations of motion are integrated
with the second order Runge-Kutta method with time step $\Delta
t=10^{-3}$.  The chosen tumbling parameters determine the value of the
single particle rotational diffusion to be $D_r=0.3$.  We employ
velocities between $0$ and $2$, such that $Pe=v / (l D_r) = 6.6$ is
the maximum value of the Peclet number.  We consider $N$ particles in
a box with dimensions $L_x$ and $L_y$, with periodic boundary
conditions in the $y$-direction, and confining walls in the $x$-direction. 
Short-range soft repulsive interactions ($V(r)\sim
r^{-12}$, with cut-off radius $r_c=3$) are considered between
the beads, and also between the beads and the walls.  The number
density is $\rho = N / A$ with $A$ the box area.  As a reference, the
density of hexagonally packed impenetrable beads of diameter $a$ is
$\rho_m\simeq 2.31$.

\begin{figure}[t]
\center\includegraphics[width=0.8\columnwidth]{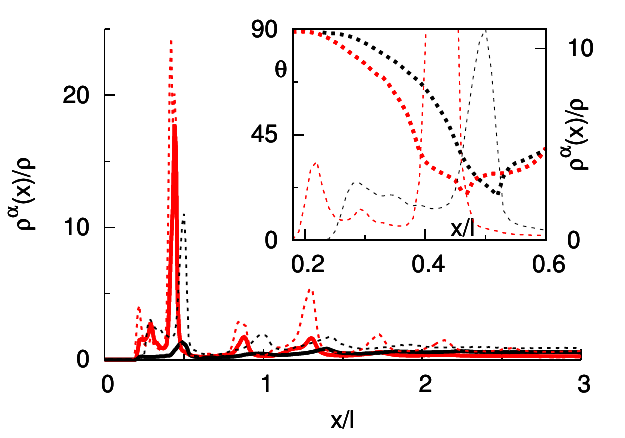}	
\caption{(Color online) Normalized local number density of the
  particles centers of mass $\rho^{\alpha}(x)$ in a microchannel as
  function of the wall distance.  Two independent one-component
  systems are displayed with dashed lines, and the two components of
  an equimolar mixture are shown with solid lines.  Black denotes
  $v=0.5$ and red $v=2$.  Inset: magnification of the region close to
    the wall.  Additional thick-dotted lines are the angle of
    particles with the wall normal vector for the one-component systems.} 
\label{fig:rho-mix}
\end{figure}

\begin{figure}[t]
\includegraphics[width=0.8\columnwidth]{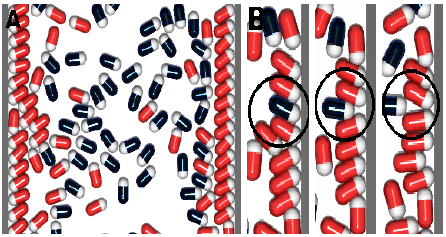}	
\caption{(Color online) Simulation snapshots of a mixture of slow
  ($v_s=0.6$, black) and fast particles ($v_f=1.4$, red).  (a)~Fast
  particles accumulate at the walls, while the slow ones are displaced
  into the bulk.  (b)~One slow particle is being {\em expelled} away
  from the wall. See also {\tt movie1.avi}. }
\label{snap1}
\end{figure}

We investigate one-component reference systems, in which all particles
have the same self-propulsion velocity $v$, and equimolar
two-component systems, in which half the particles have velocity $v_f$
and the other half velocity $v_s$, with $v_f > v_s$, and with an
average velocity $\overline{v} = (v_f+v_s)/2$.  Standard parameters
$L_y=20$, $L_x=10$, $\rho=1$, and $\overline{v}=1$ are employed unless
otherwise stated. Starting from random initial configurations, we
typically compute time averages over $10^6$ time steps
(corresponding to an average of $100$ tumbling events for a single free
particle) excluding the initial $10^5$ for equilibration, and
consider $8$ independent runs to average physical quantities of
interest.

{\bf Particle mixtures in confinement.} - First, we investigate the
properties of self-propelled particles confined in a microchannel,
with walls in the $x$-direction and periodic boundary conditions in
the $y$-direction. We are interested in characterizing the density of
particles across the channel width. We define $\rho^{\alpha}(x)$ as
the local number density of the particles centers of mass, with
$\alpha$ denoting the considered species.  In fig.~\ref{fig:rho-mix},
average normalized density profiles are displayed for two
one-component systems with velocities $v=0.5$ and $v=2$. Increasing
particle velocity enhances the accumulation of particles at the wall
(higher values of the $\rho^{\alpha}$ maximum peaks).  The
  averaged orientaion of particles with the wall normal vector, $\theta$,
  is shown in the inset of fig.~\ref{fig:rho-mix}. Motility does not
  significantly change $\theta$.  The slight peak offset in the
  density profiles is then exclusively due to the softness of the
  interaction potentials.  As a consequence of the wall accumulation,
  the density in the bulk decreases with $v$.
Figure~\ref{fig:rho-mix} also shows the density profiles of each of
the two components in a mixture.  The density peak of slow
  particles next to the walls is clearly suppressed in the mixture.
This happens via an {\em expulsion mechanism}, as shown in
fig.~\ref{snap1}b. When particles at the wall happen to be in a
configuration in which they are pushed symmetrically from both sides,
slow particles are expelled into the bulk. Therefore, fast particles
are more abundant close to the walls, and slower particles in the
bulk, as can be seen in fig.~\ref{snap1}a.

\begin{figure*}[t]
\includegraphics[width=2\columnwidth]{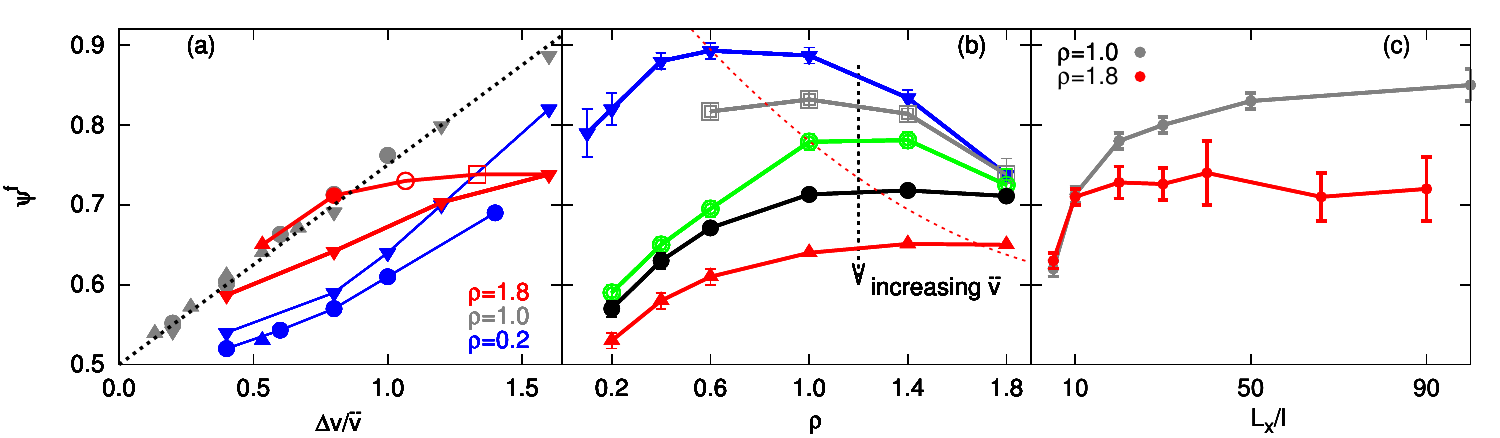}
\caption{(Color online) \label{fig:separ} Fraction of fast particles
  close to the wall, quantified by $\psi^{\alpha}$ in
  Eq.~(\ref{fraction2}). Besides standard parameters, in (b) and (c)}
  $\Delta v=0.8$. Dependence on (a) normalized velocity difference (b) density, and (c) system size. In
  (a) and (b) symbols consistently denote values of $\bar{v}$:
  $\bar{v}=0.5$ ($\blacktriangledown$), $\bar{v}=0.6$ ($\Box$),
  $\bar{v}=0.75$ ($\circ$), $\bar{v}=1.0$ ($\bullet$), $\bar{v}=1.5$
  ($\blacktriangle$).  The dashed line in (a) is the linear
  interpolation $\psi^f=0.25\Delta v/\bar{v} + 0.5$ and other lines
  are guides to the eye.  
\end{figure*}

In order to characterize the segregation of both components, we
define the separation parameter as
\begin{equation}
\psi^\alpha= \frac{\int_0^{d} \rho^\alpha(x) dx}{ \sum\limits_{\alpha}\int_0^{d} \rho^\alpha(x) dx}.  
\label{fraction2}
\end{equation}
This quantity measures the percentage of one species in the particle
population found close to the wall, indicating its relative
accumulation. We employ $d=l$, the particle length, to analyze
separation. Other distances, larger, but considerably smaller than
$L_x/2$ would provide qualitatively similar results. By construction,
$\psi^\alpha=1$ or $\psi^\alpha=0$ for complete separation, while for
completely mixed two-component system $\psi^\alpha=0.5$.

In fig.~\ref{fig:separ}, we analyze the dependence of the separation
as a function of the difference of velocities $\Delta v=v_f-v_s$,
average velocity $\bar{v}$, density $\rho$, and channel width $L_x$.
Increasing $\Delta v$ naturally leads to a larger separation.  This dependence can be understood better by
normalizing $\Delta v$ by $\bar{v}$, as shown in
fig.~\ref{fig:separ}a.  Complete separation, $\psi^f=1.0$, is expected
for a mixture with one passive component, where $\Delta v= 2\bar{v}$,
while for a system with equal velocities $\psi^f=0.5$. A linear
interpolation between both limits is then a reasonable first guess,
and it shows indeed to be a very good approximation for the behaviour
of systems with intermediate densities.  For small densities, the
accumulation at the walls is also small. The encounters of a larger
number of particles at the walls is not frequent enough for the
expulsion mechanism to be effective, therefore the separation is less
efficient.  For high densities, the accumulation at the wall is much
higher. Situations with more than one layer of particles at the wall
are frequent, resulting in a hampering of the expulsion
mechanism. Interestingly, curves for different average velocity seem
to approach nearly the same value of the separation parameter,
$\psi\simeq0.75$, for high densities. A continuous crossover between
the behavior at small and large particle densities can be observed in
fig.~\ref{fig:separ}b. The separation as a function of $\rho$ displays
a local maximum for smaller values of the average velocity $\bar{v}$,
but it is monotonically increasing with $\rho$ for larger $\bar{v}$.
The density that optimizes the separation increases with increasing
$\bar{v}$ (dashed red curve in fig.~\ref{fig:separ}b).  At constant
$\Delta v$ and $\rho$, separation decreases with increasing $\bar{v}$,
as shown in fig.~\ref{fig:separ}a and~b.  This happens because $\Delta
v/\bar{v}$ is the relevant quantity to consider, as already mentioned
above. An absolute velocity difference $\Delta v$ is less relevant
when the two velocities are high, compared to the situation when the
two velocities are low.

The channel width might also influence the separation of the two
components.  When the channel is wide enough, the accumulation at each
wall can be understood independently of the presence of the other
wall.  This is not the case for narrow channels, where particles
expelled from one wall very soon arrive at the second wall, hindering
separation.  In fig.~\ref{fig:separ}c, the separation is shown as
function of the channel width $L_x$.  The transition from weak to
  strong confinement varies with density and with the average
  free-run-length $\lambda$. At standard density $\rho=1$, we
observe a monotonic increase of the separation, reaching a plateau
value at $L_x>5\lambda$.  At high density $\rho=1.8$, the plateau
value is reached already in smaller channels $L_x \simeq \lambda$.
Other factors that influence component separation, like the particle
aspect ratio or swimming strategy, will be a matter of further
investigation.


{\bf Particle mixtures in a capillary flow.} - Flow through
microfluidic devices presents a unique possibility to study and
manipulate microswimmers.  We modify our simulation model to include
the effect of an implicit fluid with a parabolic velocity profile,
with maximum velocity $v_0$ at the center of the channel and
vanishing flow velocity at the walls.  This is performed by adding to
the translational and rotational velocities of each particle both the
velocity and the vorticity of the flow~\cite{mio-jpcm}.

\begin{figure}[t]
\center{
\includegraphics[width=0.7\columnwidth]{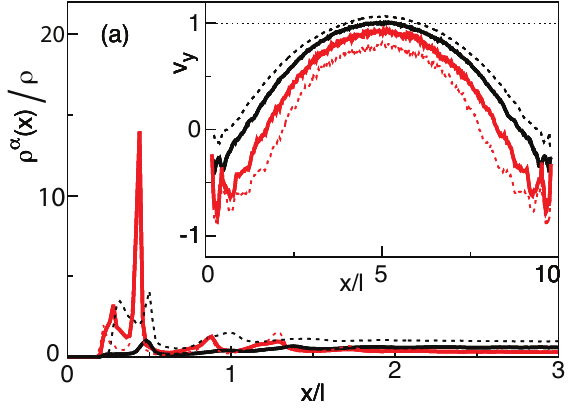}
\includegraphics[width=0.78\columnwidth]{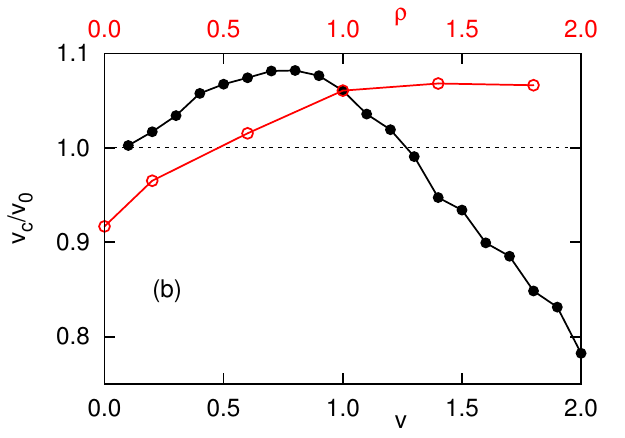}
}
\caption{(Color online) (a) Normalized local number density of the
  particles centers of mass in the presence of a capillary flow with
  $v_0=1$. Inset: velocity profiles along the channel width. Axis,
  lines, and parameters as in fig.~\ref{fig:rho-mix}. (b) Average
  velocity at the center of the channel $v_c$ as a function of
  particle density (circles) and particle self-propulsion velocity
  (bullets).}
\label{profiles-v}
\end{figure}

In the absence of flow, particles accumulate at the wall with a
  certain orientation, but always symmetric on average with respect
to the surface normal vector~\cite{mio-jpcm,Elgeti-EPL2013}.  In the
presence of flow, the interplay of self-propulsion and the vorticity
field of the flow implies that the orientation of particles
accumulated at the wall is almost exclusively opposite to the applied
flow (see fig.~\ref{profiles-v} and {\tt movie2.avi}).  This gives
rise to the interesting phenomenon of {\em upstream swimming} at the
channel walls. This affects the entire velocity profile of the
particles across the whole channel.  For a given flow field, upstream
swimming increases with self-propelling velocity.  Furthermore, the
trajectories of single self-propelled non-tumbling particles in a
parabolic flow also result in an upstream average orientation, which
enhances the upstream swimming
\cite{HillUpstream,mio-jpcm,ZoettlPRL2012,ZoettlEPJE2013}.
On the other hand, tumbling and collisions among particles randomize
the particle orientation, such that the particle average velocity
follows more easily the applied flow.  Interestingly, the velocity at
the channel center $v_c$ can become larger than the applied flow
$v_0$, see fig.~\ref{profiles-v}.  This means that the particles are
on average aligned with the flow, in contrast to the previously
described effects. This effect occurs for not too small densities, 
moderate self-propulsion velocities, and it is most likely related to
a cooperative phenomenon.  The combination of these effects can result
in the non-monotonic dependence of the velocity at the channel center
$v_c$ with the self-propelling velocity, as shown in
fig~\ref{profiles-v}b, where the increase of $v_c$ with density is
also displayed.

The question that arises now is how the presence of flow affects the
spontaneous separation in mixtures of self-propelled particles with
different velocities.  The local particle density $\rho^\alpha(x)$,
shown in fig.~\ref{profiles-v}, indicates that the fast particles
still expel the slow ones from the near-wall region into the bulk, and
that the presence of flow slightly diminishes layer formation in the
proximity of the walls, similar to the one-component
case~\cite{mio-jpcm}.  The velocity profiles of each component
$v^\alpha_y(x)$ are displayed in fig.~\ref{profiles-v}.  Both profiles
differ from their counterparts in the one-component systems by getting
closer to each other, which reflects the effective drag of one
particle component on the other.

\begin{figure}[t]
\center\includegraphics[width=0.85\columnwidth]{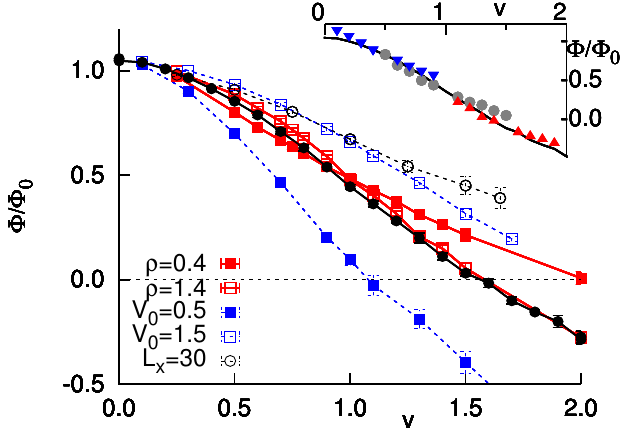}
\caption{(Color online) Normalized flux of one-component systems as a
  function of the self-propelling velocity. Labels indicate
  non-standard parameters. Inset: the line corresponds to the
  one-component reference flux, and symbols to the mixture components.
  Four mixtures with different $\Delta v$ are displayed for each
  $\bar{v}$.  Symbols indicate $\bar{v}=1.5$~($\blacktriangle$),
  $\bar{v}=1.0$~($\bullet$), and $\bar{v}=0.5$~($\blacktriangledown$).
  Fluxes are normalized such that \mbox{$\Delta v\rightarrow 0
    \Rightarrow \phi^\alpha\rightarrow\phi$}.  }
\label{fig:no-mix}
\end{figure}

In order to quantify the separation of components in the presence of
flow, we compute the particle flux $\phi^\alpha$ of each
component, which is defined as 
\begin{equation}
\phi^\alpha=\int_0^{L_x}\rho^\alpha(x) v_{y}^\alpha(x) dx.
\label{eq:flux}
\end{equation}
This quantity expresses the number of particles crossing a channel
section in a unit of time, and its value is a consequence of the
interplay of density and velocity profiles.  In the case of passive
tracer particles with a flat density profile moving in a parabolic
flow with maximum velocity $v_0$, the flux can be calculated easily to
be $\phi_0=2\rho v_0 L_x/3$.  In fig.~\ref{fig:no-mix}, the normalized
flux $\phi/\phi_0$ is displayed for various one-component systems as a
function of $v$. The limit of passive particles $\phi\to\phi_0$ is
reasonably well reproduced, given that the density profile is never
perfectly flat in the presence of confining walls.  With increasing
self-propulsion velocity, the flux decreases as a consequence of the
increase of the upstream swimming.  In general, however, the
dependence of the normalized flux on other system parameters is
non-trivial. Increasing the average density shows a non-monotonic
albeit very small variation of the normalized flux.  Increasing the
fluid velocity increases the normalized flux, which indicates that the
particles follow the flow more the stronger it is, which is also the
effect of enlarging the channel width.

In a mixture, the changes in the velocity profiles translate into a
slight increase of the fast-particle flux relative to the single-component case, 
while for slow particles the flux slightly decreases.
This is shown in the inset of fig.~\ref{fig:no-mix}, where the
one-component reference normalized flux is compared with the values
for each of the mixed components.  Only for $\bar{v}=0.5$, we observe
a slight flux increase for both components.  In this case, the
interplay between density and velocity is different, and the
separation is more effective.

\begin{figure}[t]
\center\includegraphics[width=0.8\columnwidth]{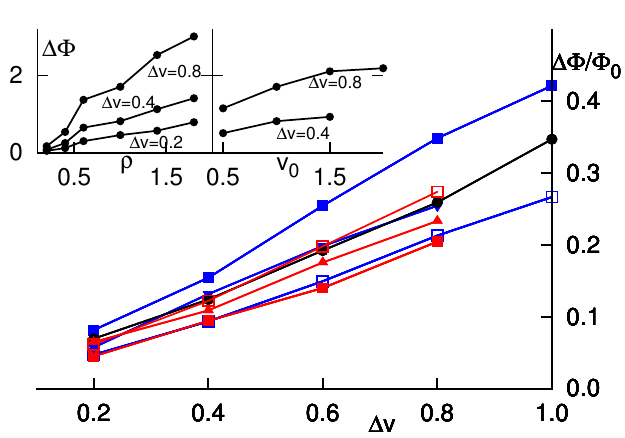}
\caption{(Color online) 
  Normalized flux difference for two-component systems, as a function 
  of $\Delta v$. Inset: flux difference as a function of density and fluid 
  velocity.  Symbols are the same as in fig.~\ref{fig:no-mix}. 
}
\label{fig:flux}
\end{figure}

The separation of fast and slow particles obtained in the outflow can
be characterized by the difference of the fluxes of both components
$\Delta\phi=\phi^s-\phi^f$. This difference increases with density,
fluid velocity, and channel width, similar to the normalizing flux
$\phi_0$, and naturally also with the difference of velocities, as
displayed in the inset of fig.~\ref{fig:flux}.
Figure~\ref{fig:flux} shows that the largest normalized flux
differences occur for the smallest employed flow velocity.  At that
flow velocity the single-component normalized flux shows the largest
decrease with self-propulsion velocity (fig.~\ref{fig:no-mix}).  As we
have shown above, the mixing only slightly changes the flux values,
such that the behavior of the single-component fluxes can be taken as
a reference for flux differences in mixtures.  An interesting
consequence arising from the flux presented in
fig.~\ref{fig:no-mix} is that for a given mixture, the applied flow
velocity can be eventually tuned to obtain a positive flux for the
slow particles and a negative flux for the fast particles. For
example, this can be achieved for a mixture with $v_s=0.5$ and
$v_f=1.5$ and standard parameters by choosing $v_0=0.5$.

\begin{figure}
\includegraphics[width=0.75\columnwidth,height=4cm]{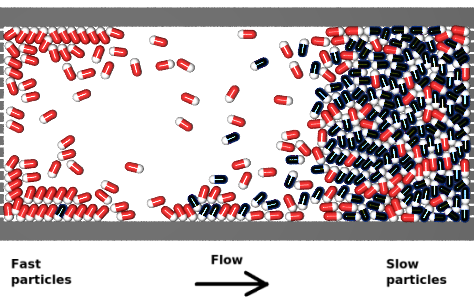}
\center
\includegraphics[width=0.7\columnwidth]{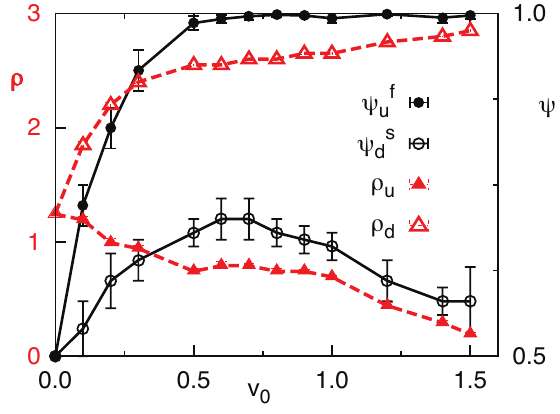}	
\caption{(Color online) (a) Simulation snapshot of a micro-channel
with capillary flow ($v_0=1.0$) and confining membranes.
Fast particles ($v_f=1.5$) are found upstream and slow ones
($v_s=0.5$) downstream. (b) Percentage of fast particles in
the upstream quarter of the channel $\psi^f_u$, and of slow
particles in the downstream quarter $\psi^s_d$
as a function of $v_0$. Total
number densities in each quarters, $\rho_u$ and $\rho_d$.}
\label{fig:membrane}
\end{figure}

{\bf Particles mixtures in microchannel flow confined between
  membranes.} - An interesting configuration to exploit the separation
of self-propelled particles of different self-propulsion velocities in
microcapillary flow is shown in fig.~\ref{fig:membrane}a. It consists
of two semi-permeable membranes spanned across the channel diameter at
a defined channel segment.  The membrane pores are assumed to be too
small for the particles to penetrate, but large enough for the fluid
flow to pass unhindered. Thus, in our simulations, the fluid flow is
unperturbed compared to the open channel, while excluded-volume
interactions are implemented for the self-propelled particles.

At the walls, the particles mostly move fast and upstream, which in
the presence of the membranes leads to the accumulation of almost
exclusively fast particles at the upstream membrane, as shown in
fig.~\ref{fig:membrane}a and in {\tt movie3.avi}. In the center of the
channel, equidistant from the walls, both particle types are present
and move downstream, but slow particles are more abundant. This
leads to an accumulation of a mixture of particles that is enriched in
slow particles at the downstream membrane.  The competition of these
two effects determines the precise distribution of particles along the
channel, which therefore strongly depends on the system parameters.
The intrinsic distribution of self-propelled particles in the channel
also translates in the accumulation at the membranes, which is much
more ordered upstream than downstream.

In order to quantify the separation, we compute the percentage of fast
particles at a distance $d$ from the upstream-membrane,
$\psi^f_u$, and the percentage of slow particles at a distance $d$
from the downstream-membrane, $\psi^s_d$, which is calculated using
Eq.~(\ref{fraction2}) with $d=L_y/4$.  Relevant quantities are
also the total particle densities $\rho_u$ and $\rho_d$ in both regions.
The dependence of the separation parameters and densities close to the
membranes on the applied flow velocity is presented in
fig.~\ref{fig:membrane}b for a mixture with $v_s=0.5$ and $v_f=1.5$.
In the absence of flow, there is no separation, $\psi^\alpha=0.5$, and
both densities are equal to $\rho=N/A$, which in this case is $1.25$.
By increasing the fluid velocity, the particles are increasingly
pushed to accumulate at the downstream membrane, reaching a packing
limit where neighboring particles can be slightly closer than the bead
diameter due to the soft interaction potential.  Interestingly, only
fast particles are found at the upstream membrane for $v_0\geq v_s$,
while the largest proportion of slow particles is found at the
downstream membrane for flow velocities a bit larger than $v_s$.

{\bf Conclusions.} - The separation of run-and-tumble particles in
terms of their motility in a microchannel has been studied.  Slow
particles are shown to be {\em expelled} by the fast neighboring
particles from the wall into the bulk, which results in the
spontaneous segregation of the two components.  This mechanism can be
of importance to achieve particle sorting, {\em e.g.},  via a pipette
or a microfluidic device.  We have also investigated how the presence
of a capillary flow in the microchannels affects segregation.  In this
case, since particles with different velocities have different fluxes,
the fluid velocity can be tuned to segregate slow particles downstream
and fast particles upstream.  Separation always grows with the
velocity difference, but the dependence on other parameters is less
obvious.  In the absence of flow, separation is maximized for low
average particle velocities and intermediate densities. In the
presence of flow, separation grows with fluid velocity, channel width,
and density, while there is almost no dependence on the average
particle velocity.  Moreover, the separation of fast and slow
particles in a channel with flow and membranes at its ends has shown to
be the maximal for fluid velocities slightly higher than the slow-particle velocity.
 In order to ensure the generalization of our
  conclusions beyond the details of our model, we have performed
  additional simulations which show that the expulsion mechanism also occurs
  in mixtures of spherical particles, other swimming strategies
like active Brownian particles, and in other confining
geometries as the presence of obstacles.  Qualitatively, the 
  results are expected to persist in three dimensional structures, as is
  the case for a single swimmer in Poiseuille flow~\cite{ZoettlEPJE2013}.

Our results should be of considerable
theoretical and practical interest, because they provide new insights
into active matter and non-equilibrium systems.  Applications of these
results can be envisioned for the development of multi-stage cascade
microfluidic lab-on-chip devices, which could then be an alternative
to the use of laser traps or other external fields, for sorting
particles with different motilities.

We thank L. Angelani, R. Di Leonardo and A. Wysocki for very
useful discussions.  A.C. acknowledges partial support by the
International Helmholtz Research School of Biophysics and Soft Matter
(IHRS BioSoft).

\end{document}